\documentclass{aastex}
\usepackage{emulateapj5}
\shorttitle{A Chandra view of Naked AGN}
\shortauthors{Gliozzi et al.}

  \def\chandra{{\it Chandra}} 
  \def\xmm{{\it XMM-Newton}}

  \def\lum{erg s$^{-1}$}

  \def\arcsec{$^{\prime\prime}$}

  \def\ltsima{$\; \buildrel < \over \sim \;$}
  \def\simlt{\lower.5ex\hbox{\ltsima}} 
  \def\gtsima{$\; \buildrel > \over \sim \;$}
  \def\simgt{\lower.5ex\hbox{\gtsima}} 

\begin{document}
\title{A Chandra view of Naked Active Galactic Nuclei}

\author{Mario Gliozzi}
\affil{George Mason University, 4400 University Drive, Fairfax, VA 22030}

\author{Rita M. Sambruna}
\affil{NASA's Goddard Space Flight Center, Code 661, Greenbelt, MD 20771}

\author{Luigi Foschini}
\affil{INAF/IASF-Bologna, via Gobetti 101, 40129 Bologna, Italy}

\begin{abstract}
We present the first X-ray observations of three sources belonging to a 
new AGN class: the naked AGNs. Based on optical spectroscopic studies, these 
sources appear as classical type 2 (obscured) AGNs, with only narrow emission 
lines.  However, long-term optical monitoring campaigns, carried out over more 
than two decades, show that the same sources are strongly variable, like 
type 1 (un-obscured) AGNs. Based on short Chandra observations, 
the sources appear to be fairly bright in the X-rays, with typical Seyfert 1s
values for the photon index ($\Gamma\sim$1.8) and without significant 
intrinsic 
absorption, supporting the conclusion that some bright AGNs may genuinely lack 
a broad line region. Future, broad-band studies as well as
deeper X-ray observations, probing both the 
spectral and the temporal properties of the naked AGNs, are crucial to shed 
light on the central engine of these sources, which may be representative of 
a large class of AGNs.
\end{abstract}


\section{Introduction}
More than two decades of multi-wavelength observations and theoretical studies
have led to a generally accepted unification scheme for AGNs. According to this
scheme, all AGNs share some basic ingredients: a supermassive black hole; 
an accretion disk coupled with a hot corona,
which radiates from the optical through X-ray energies;
high velocity and high density ($n_{\rm e} > 10^9~{\rm cm^{-3}}$) gas located 
at $\sim$ pc scales, usually referred to as the broad-line
region (BLR), since its radiation is dominated by broad (velocity dispersion 
of $\sim$ 2000--10000 km ${\rm s^{-1}}$) permitted lines; lower velocity and 
low density ($n_{\rm e} = 10^3-10^6~{\rm cm^{-3}}$) gas located at $\sim$ kpc 
scales, referred to as the narrow-line region (NLR), and emitting narrow (velocity
dispersion $<$ 1000  km ${\rm s^{-1}}$) forbidden and permitted lines; a 
gaseous and molecular torus on pc scales presumably obscuring the inner engine
and the BLR when viewed edge-on; and, in $\sim$ 10\% of the AGNs (the 
radio-loud AGNs), a relativistic jet extending up to Mpc scales, whose
axis is generally assumed to be parallel to the torus axis. Within
this scheme, several observational differences in the broad-band
and line spectral properties between type 1 and type 2 AGNs are explained by
orientation effects (Urry \& Padovani 1995; Antonucci 1993).

Although observations generally support orientation-based unification models
for AGNs, exceptions do exist. For example, only about 50\% of the brightest 
Seyfert 2s show the presence of  hidden broad line regions (HBLRs; Tran 2001) 
in their polarized optical spectra. Several studies indicate 
that the presence or absence of HBLRs depends on the AGN luminosity, with HBLR
sources having larger luminosities (e.g., Tran 2001; Gu \& Huang 2002). More 
specifically, Nicastro (2000) hypothesized that the absence of HBLRs 
corresponds to low values of accretion rate onto the central black hole. X-ray 
observations, which are less affected by absorption compared to optical 
studies,
indicate that, in several cases, the 
non-detection of a BLR cannot be ascribed to obscuration both in radio-quiet 
(e.g., Georgantopoulos \& Zezas 2003; Nicastro et al. 2003) and in radio-loud 
AGNs (e.g. Gliozzi et al. 2004). However, for column densities
$N_{\rm H} > 10^{24}{~\rm cm^{-2}}$, i.e. for Compton thick AGNs, also the 
X-rays
at energies $<$ 10 keV are completely absorbed. Therefore, in those cases
 neither \chandra\ 
nor \xmm\ can measure the intrinsic absorption or
probe the nuclear region directly (see Maiolino \& Risaliti
2007 for a recent review).

Recently, Hawkins (2004) reported the discovery of a new class of AGNs,
apparently characterized by the absence of the broad line region 
accompanied by strong continuum emission and strong variability in the optical
band. This result is based on a large-scale monitoring program carried out in 
different optical bands over a period of 25 years, and on the spectroscopic 
follow-up (see Hawkins 2004 and references therein for a more detailed 
description of the optical observations). Of the 200,000 objects observed 1500
were selected as AGN candidates. The sample was further reduced to 55 
Seyfert-like candidates, after measuring the redshift of all emission line 
objects and selecting only sources with an adequate signal-to-noise ratio. 

An effective way to discriminate between the various classes of AGNs is based 
on the width of the Balmer lines and on the [O III]$\lambda$5007/H$\beta$ line
ratio (see Figure 5 of Hawkins 2004). Type 1 AGNs are characterized 
by H$\beta$ 
FWHM  $>2000{~\rm km~s^{-1}}$ and [O III]$\lambda$5007/H$\beta< 3$. On the 
other hand, type 2 AGNs typically show H$\beta$ FWHM  $<1000{~\rm km~s^{-1}}$ 
and large values of [O III]$\lambda$5007/H$\beta$. 
Interestingly, Hawkins finds that six objects with high 
[O III]$\lambda$5007/H$\beta$ line ratios (i.e., type 2 AGNs) show a strong 
variability typical of type 1 AGNs. These are the {\it naked AGNs}.
Note that, based on their optical properties, the naked AGNs cannot be 
classified as narrow line Seyfert 1 galaxies (NLS1), which are characterized 
by H$\beta$ FWHM  $<2000{~\rm km~s^{-1}}$,  
[O III]$\lambda$5007/H$\beta< 3$ and strong FeII (see Pogge 2000 for a review).

The existence of AGNs without BLR is apparently at odds with the standard
unification models. However, it must be outlined that, although the
unification models are able to explain most of the AGN observational
properties, they do not offer any insight into the physical origin of some
basic AGN ingredients, namely the BLR, the NLR, and the torus. Therefore, 
detailed investigations of AGNs without BLR have the potential of shedding
some light on the physical mechanisms that lead to the suppression 
(and possibly also to the formation) of the BLR.

Previous detections of AGNs without BLR were mostly based on the 
spectro-polarimetric observations, which require data with very high 
signal-to-noise (S/N) ratio
and rely on the existence of an appropriately placed scattering region
to view the obscured nucleus. Therefore, hidden BLR can be missed by this
method, leading to misclassified AGNs without BLR. On the other hand, the
detection of naked AGNs relies on the combined use of temporal and spectral 
properties. In principle, the former ensures that the nucleus is viewed 
directly (via the long-term variability typical of type 1 AGNs), 
whereas the latter confirms
the existence of a typical NLR excluding the presence of a BLR. 

However,
Hawkins (2004) does not explicitly address the important issue 
of the source of the observed optical variability. If the latter
is related to an unresolved jet component, we cannot a priori exclude that
the AGN nuclear region (including the BLR) is obscured and that
the naked AGNs are not ``pure'' type 2 objects. In addition,
recent studies have shown that the optical classification
of type 1 /type 2 objects is not trivially related to the classification
obtained via X-ray observations (e.g., Bauer et al. 2004; Cappi et al. 2006).
An investigation of the broad-band properties is clearly necessary to
confirm the genuine lack of BLR and hence shed some light on the nature 
of these enigmatic AGNs.

One of the most effective ways to investigate the properties of AGNs is with 
X-ray
observations, since: 1) The X-rays are produced and reprocessed in the inner, 
hottest nuclear regions of the source. 2) The penetrating power of (hard) 
X-rays allows 
them to carry information from the inner core regions without being 
substantially
affected by absorption, and to probe the presence of the putative torus. 3) 
Compared
to optical radiation, X-rays are far less affected by the host galaxy 
contamination.
Interestingly, the six naked AGNs, which have been extensively studied at 
optical wavelengths, have never been detected in the X-ray band, either 
because dedicated pointed X-ray observations have never been performed, or 
because X-ray sky surveys performed by previous satellites were not sensitive
enough to detect them. 

Here, we report the first X-ray study of three naked AGNs obtained with 
\chandra\ snapshot observations. The primary goal of this project is to 
verify whether the
sources are detected at X-ray energies and derive the basic information
(count rate and spectral slope) necessary to refine their X-ray investigation 
with deeper observations. In addition, we use the X-ray properties, 
complemented with information from the optical band, to investigate the
possible presence of a jet and to compare the naked AGNs with
other AGNs without BLR obtained with spectro-polarimetric studies.

The outline of the paper is as follows. In $\S~2$ we describe the observations
and  data reduction. The main findings are reported in $\S~3$. In $\S~4$ we
discuss the results, and  $\S~5$ we summarize the main conclusions.
Hereafter, we adopt $H_0=71{\rm~km~s^{-1}~Mpc^{-1}}$, $\Omega_\Lambda=0.73$ and
$\Omega_{\rm M}=0.27$ (Bennet et al. 2003)

\section{Observations and Data Reduction}
Three of the naked AGNs discovered by Hawkins,
Q2122-444, Q2130-431, Q2131-427, were observed with \chandra\ ACIS-S in 
December 2005. Since we had no X-ray 
information on these sources, for precaution the observations were carried 
out  using subarray modes (with frame time ranging between 0.4 and 0.8 s) to
minimize possible pile-up effects.
The general source properties from Hawkins (2004) are summarized in 
Table 1. The values for the broad-band spectral index $ \alpha_{\rm OX}$,
defined as $0.3838 
\log(l_{\rm 2keV}/l_{\rm 2500\AA})$, have been obtained from the best-fit
parameters of the $ \alpha_{\rm OX} - l_{\rm 2500\AA}$ relation
provided by Steffen et al. (2006). The B magnitudes  have been converted into 
fluxes at
2500\AA\ assuming a typical optical slope of 0.7 ($F_\nu\propto \nu^{-0.7}$).

The data, provided by
Chandra X-Ray Center, were processed  using \verb+CIAO+ v. 3.2.2
and \verb+CALDB+ v. 3.2 following standard criteria. 
Only events with 
grades 0, 2--4, and 6 and in the energy range 0.3--8 keV were retained. 
We also checked that no flaring background events occurred during the 
observations. The extraction
of images and spectra, as well as the construction of response matrices,
were obtained following the standard procedure. 
Background spectra and light curves were extracted from source-free regions 
on the same chip of the source.
The observation log with
dates of the observations, net exposures, frame times, counts, and the 
average count rates are reported in Table 2. The counts were extracted from
circular regions with radius of 2\arcsec\ centered on the nominal optical
positions. The uncertainties
for the X-ray counts were
calculated following Gehrels (1986), as appropriate in a regime of low
counts.

The spectral analysis  was performed using the {\tt XSPEC v.12.3.0}
software package (Arnaud 1996; Dorman \& Arnaud 2001). 
Given the low number of counts in each
spectrum, due to the short exposure and the moderate count rate, we 
used the $C-$statistic (Cash 1979)
to fit the un-binned spectra over an energy range of 
0.5--8 keV (in the observer's reference frame), where the calibration is best 
known (Marshall et al. 2005) and the background negligible.
The errors on spectral parameters are at 90\% confidence level for one 
interesting parameter ($\Delta C=2.71$).

\section{X-ray Results}

\subsection{Imaging}
\begin{figure}     
 \includegraphics[bb=5 180 609 614,clip=,angle=0,width=6.cm]{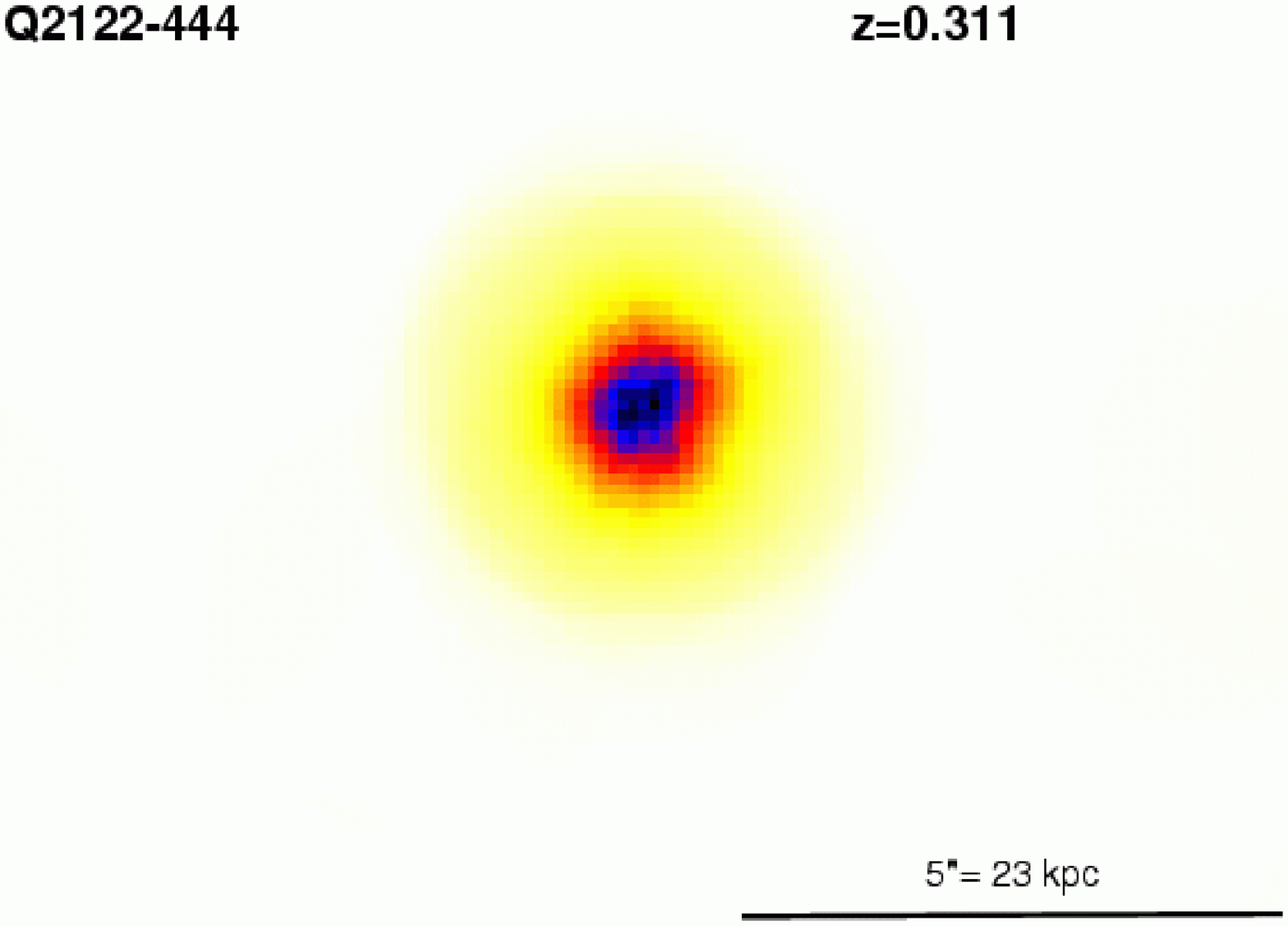}\includegraphics[bb=5 180 609 614,clip=,angle=0,width=6.cm]{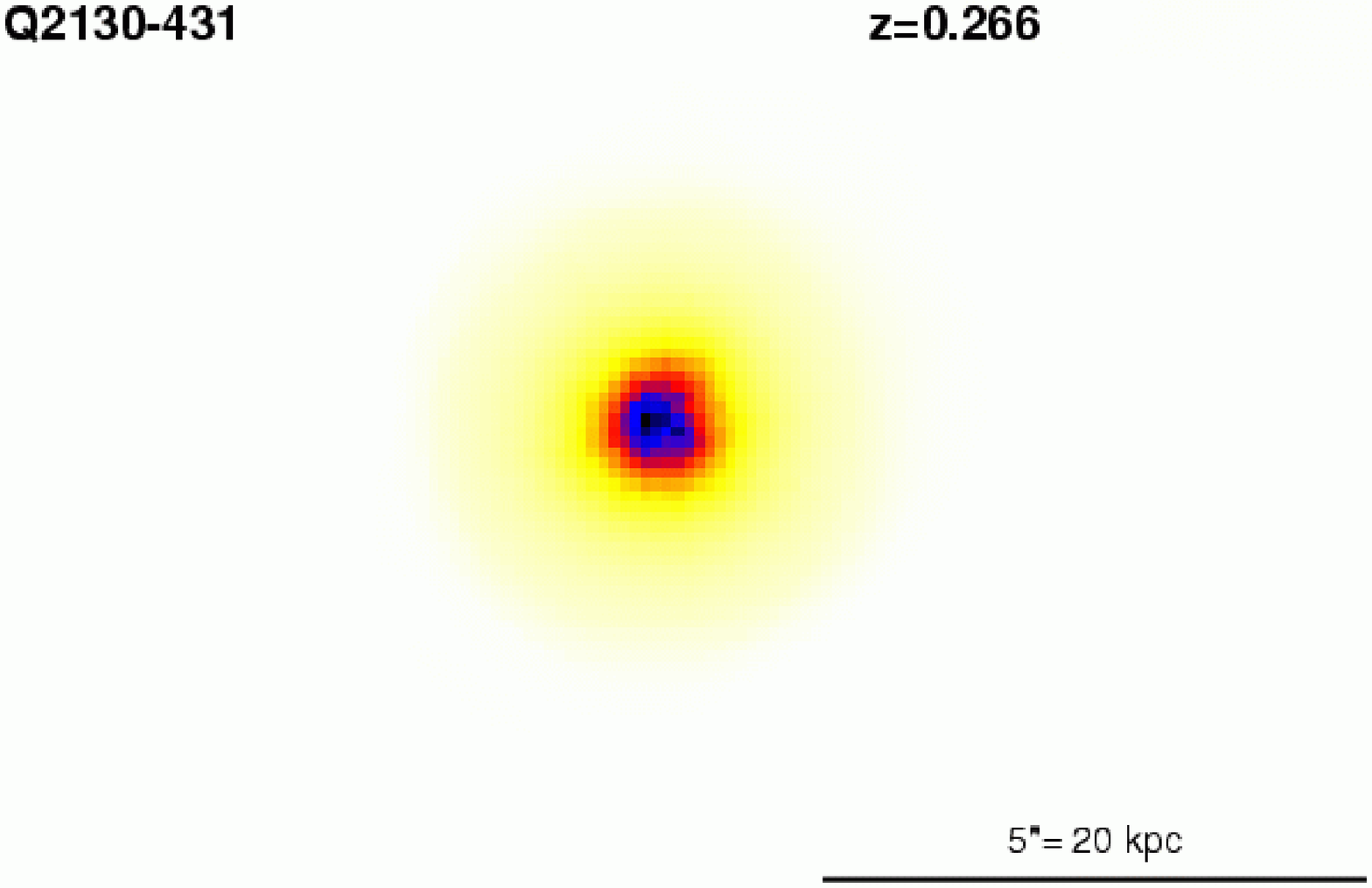}\includegraphics[bb=5 180 609 614,clip=,angle=0,width=6.cm]{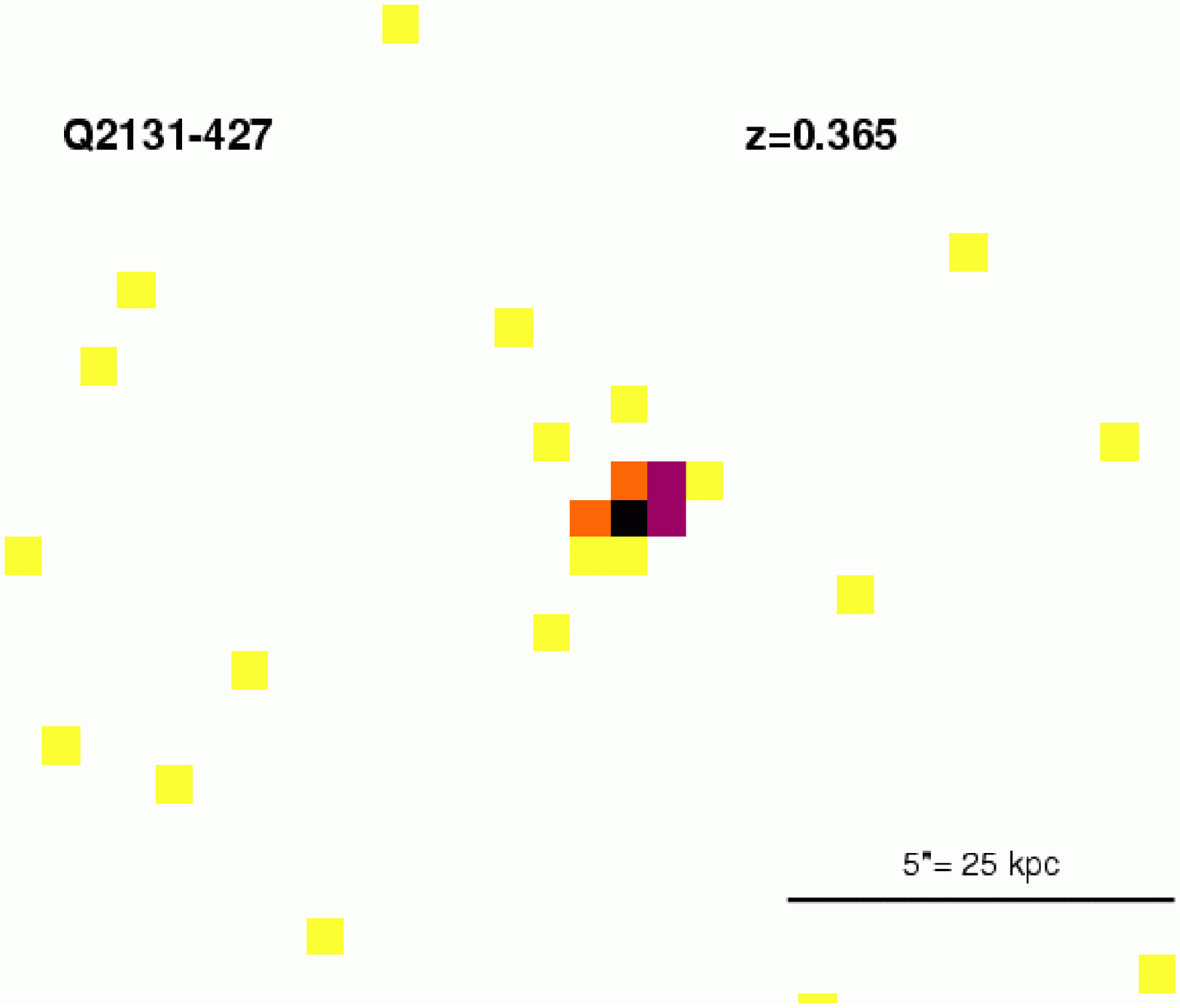}          
\caption{\chandra\ ACIS-S  images of Q2122-444, Q2130-431, Q2131-427 in the 0.3--8~keV band.
The left and middle images were smoothed using the sub-package {\it fadapt} of     
{\it FTOOLS} with a circular top hat filter of adaptive size, adjusted so as to
include a minimum number of 10 counts under the filter; each final     
pixel is 0{\farcs}1. The image of  Q2131-427 (right panel) is unsmoothed.}  
\label{figure:fig1}   
\end{figure}

Figure~\ref{figure:fig1} show the \chandra\ images of the three naked AGNs 
in the total energy range 0.3--8 keV. In order to fully exploit the 
unsurpassed spatial resolution of \chandra, for the two brightest sources
Q2122-444 and Q2130-431 (left and middle panels) the images were
extracted from ACIS-S event files after rebinning with a factor of 0.2 
(yielding pixel size of 0.1\arcsec) and smoothing with the subpackage 
\verb+fadapt+ of \verb+FTOOLS+, with a circular top hat filter of adaptive 
size in order to achieve a minimal number of 10 counts under the filter.
This procedure cannot be applied to Q2131-427 for the low number of counts.
As a consequence, for this source we simply extracted a
raw image, which is  shown in the right panel of Fig~\ref{figure:fig1}.
All the sources are clearly detected during the short \chandra\ exposures,
and no nearby objects are present within a radius of 1\arcmin\ from the central
AGN. 

\chandra, with its sub-arcsecond spatial resolution, is the best X-ray
satellite to investigate the presence of extended emission, which is of great
interest in AGN studies for the following reasons. 
Firstly, if the bulk of the X-rays 
arises from an extended region rather than from a compact nuclear source,
the X-ray emission is likely to be dominated by starburst activity  and not 
by an AGN (e.g., Dudik et al. 2005). Secondly, the presence of 
circum-nuclear diffuse X-ray emission
in low redshift AGNs, might be used to probe the AGN fuel reservoir and thus to
put constraints on the radiative efficiency (e.g., Gliozzi et al. 2003 and
references therein). Finally, and more importantly for the naked AGNs, an
asymmetric departure from the azimuthally symmetric point spread function (PSF)
model may reveal the presence of X-ray jets, which have been frequently 
detected in
longer \chandra\ observations (e.g.  Sambruna et al. 2002).

The visual inspection of Fig.~\ref{figure:fig1} suggests that
the three AGNs are point-like sources at the \chandra\ spatial resolution.
This conclusion is confirmed by the analysis of their radial profiles, which 
can be adequately fitted by a PSF plus background.
The PSF was created using the \chandra\ Ray Tracer (\verb+ChaRT+) simulator 
which takes into account the spectrum of the source and its location on the 
CCD. However, the short exposures hamper an accurate spatial analysis and 
thus the presence of a weak extended emission cannot be ruled out based on 
the current data.

\subsection{Spectral analysis} 
\begin{figure}
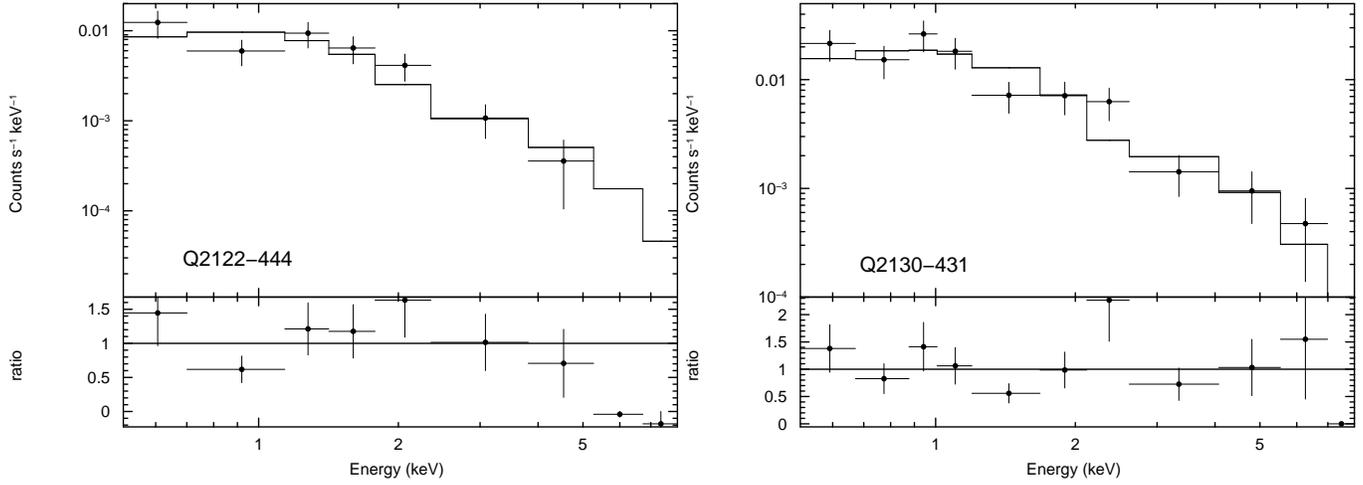

\includegraphics[bb=40 2 567 704,clip=,angle=-90,width=9.cm]{f2a.eps}\includegraphics[bb=40 2 567 704,clip=,angle=-90,width=9.cm]{f2b.eps}
\caption{Spectra of Q2122-444 and Q2130-431 and data/model ratios to a simple power-law model modified by
photoelectric absorption in the Milky Way.}
\label{figure:fig2}
\end{figure}

\begin{figure}
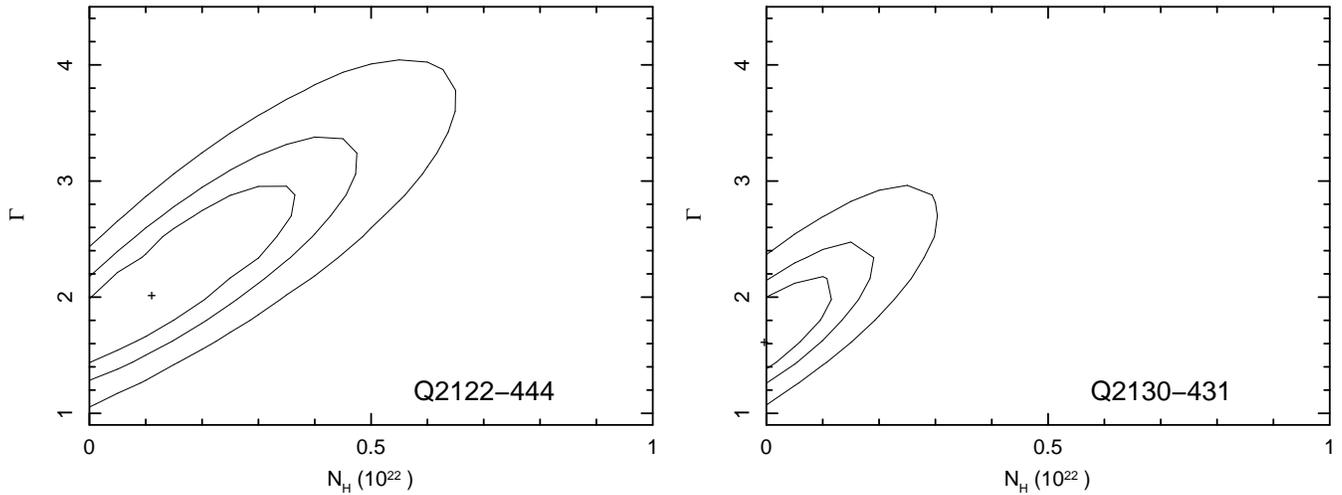

\includegraphics[bb=120 95 545 640,clip=,angle=-90,width=9.cm]{f3a.eps}\includegraphics[bb=120 95 545 640,clip=,angle=-90,width=9.cm]{f3b.eps}
\caption{Confidence contours (68\%, 90\%, and 99\%) in the $\Gamma$ --
$N_{\rm H}$ plane for Q2122-444 and Q2130-431.}
\label{figure:fig3}
\end{figure}

Two of the three AGNs observed with \chandra, Q2122-444 and Q2130-431, have 
enough counts to allow a direct fitting of the un-binned spectrum using the 
$C-$statistic. Both sources are reasonably well fitted with a simple 
power-law model absorbed by Galactic $N_{\rm H}$. The values of the photon 
indices  ($\Gamma=1.8\pm0.4$ for both Q2122-444 and  Q2130-431) are typical 
for Seyfert 1 galaxies (e.g., Nandra et al. 1997), although with substantial 
uncertainties due to the limited photon statistics.
The spectra and residuals of Q2122-444 and Q2130-431 are shown in 
Figure~\ref{figure:fig2}. If $N_{\rm H}$ is left free to vary in the spectral 
fits, the resulting spectral parameters are  poorly constrained, as outlined 
by  Figure~\ref{figure:fig3} that shows the confidence contours of $\Gamma$ 
and  $N_{\rm H}$.
On the one hand, these results indicate that snapshot \chandra\ observations 
are inadequate for a detailed spectral analysis. On the other hand, the low 
upper limits obtained for $N_{\rm H}$ suggest that the sources are not heavily
absorbed (provided that the bulk of the X-rays is indeed produced by the AGN 
and not by the jet).

The third source, Q2131-427, has less than 20 counts in the 0.5--8 keV energy 
band, which are not enough for a meaningful spectral fitting: The spectrum can
be fitted with a simple power law, but the photon index, $\Gamma=2.1\pm0.9$, 
is basically unconstrained. Nevertheless, the spectral properties of Q2131-427
can be probed by computing the hardness ratio $HR=(h-s)/(h+s)=-0.60\pm0.27$ 
(where $s$ refers to the 0.5--2 keV energy band and $h$ to 2--8 keV).

The photon indices,  obtained by fixing $N_{\rm H}$ to the Galactic value,
along with the hardness ratios, the observed fluxes, and the intrinsic 
luminosities are summarized in Table 3. Since the uncertainties are dominated 
by the errors on the photon index, the errors on fluxes and luminosities have 
been derived by varying $\Gamma$ by $\pm1\sigma$.
The similarity of the hardness ratio of Q2131-427 with the $HR$ values 
obtained for Q2122-444 and Q2130-431 (see Table 3) suggests that also the 
spectrum of Q2131-427  is consistent with a simple power law with a typical 
Seyfert-like photon index.

For completeness, in the last column of Table 3, we have also reported
the 2 keV monochromatic luminosities expected on the basis of the 
$\alpha_{\rm OX}$ values, which in turn  were obtained
 assuming the best-fit parameters of the correlation between
$\alpha_{\rm OX}$ and $l_{2500\AA}$  proposed by Steffen et al. (2006).
The measured X-ray luminosities are
roughly consistent with the predicted values, with the values for Q2122-444 
and Q2131-427 lower by a factor 1.5--2 compared to the expected values, and
with Q2130-431 higher by a factor of 1.5.
This discrepancy can be explained by the fact that the X-ray
and optical measurements are not simultaneous and these sources are
known to be highly variable. Alternatively, it
might indicate that not only the line properties but 
also the broad-band 
continuum of the naked AGNs is different from that of more typical AGNs.

\section{Discussion}
One of the key property of the naked AGNs is their pronounced and 
continuous variability observed in the optical band over a period of 25 years.
This might indicate that the central region of the naked AGNs is viewed 
directly and thus that lack of the BLR cannot be ascribed to low-quality 
observations or to obscuration effects. On the other hand, this variability
might be a signature for an unresolved jet component, casting doubts on the
conclusion that the naked AGNs are pure type 2 objects.

Although a jet dominance in the continuum properties of the naked AGNs
cannot be ruled out a priori, we regard this possibility as unlikely for
several reasons. First of all, the jet dominance is observed in blazars,
which represent a very small fraction of the AGN population ($<1\%$), 
while Hawkins found that more than 10\% of the monitored Seyfert-like objects
have significant variability. Secondly,
a direct comparison of the naked AGN light curves with those typically
observed in blazars reveals several differences.
Compared to jet dominated sources (e.g., Raiteri et al. 2006),
the naked AGNs seem to be characterized by smoother, lower-amplitude 
variability without short-term events. Furthermore, the strength of the 
optical lines typically observed in jet-dominated AGNs is weaker than
that observed in naked AGNs (cf, Sbarufatti et al. 2006). Finally,
the 2--10 keV luminosity measured for naked AGNs is systematically lower than
the  values obtained for blazars that show optical lines (Donato et al.
2005). 

Although the above suggestive evidences seem to indicate that the
continuum emission in naked AGNs is not jet-dominated, only a broad-band
investigation including infrared and radio data will be able to 
definitively rule out the presence of a jet in naked AGNs. We plan to carry out
this kind of analysis in the near future, along with deeper X-ray 
observations. 
The latter will focus at investigating not only the time-averaged
spectra, but also the temporal and spectral variability properties, which 
might be crucial to discriminate between radio-loud and radio-quiet 
scenarios (see, e.g., Gliozzi et al 2006). For the rest of the discussion, 
we will assume that the bulk of the nuclear emission is associated with
the accretion process.

It is instructive to compare the properties of naked AGNs with those of
other AGNs lacking a BLR discovered with spectro-polarimetric observations.
For example, the median value of [O III]$\lambda$5007/H$\beta$ derived by Tran
(2001) for non-HBLR Seyfert 2 galaxies (the pure Seyfert 2s) -- $6.8\pm1.5$ --
is fairly consistent with the ratio measured for Q2122-444 and Q2131-427, but
much lower than the value obtained for Q2130-431 (see Table 1). The other 
properties discussed by Tran (2001) are related to the infrared band and thus 
cannot be used for a direct comparison with the naked AGNs, which only have 
optical and X-ray observations. 

On the other hand, Nicastro et al. (2003) discuss the X-ray properties
of sub-sample of non-HBLR Seyfert 2s obtained by Tran (2001). The values for 
the 2--10 keV luminosity from Nicastro et al. (2003) are significantly lower 
than the luminosities measured by \chandra\ for the 3 naked AGNs. However, 
according to Nicastro et al. (2003), the driving parameter is the accretion 
rate (in Eddington units), not the X-ray luminosity. Therefore,  in order to 
carry out a direct comparison with the findings of the above authors, we need 
to estimate the black hole masses of the naked AGNs. This can be done by 
exploiting the measurement of FWHM([O III]) (Green \& Ho 2005). However, it 
must be kept in mind that this is probably not a robust indicator (Netzer \& 
Trakhtenbrot 2007) and thus the values for the black hole 
mass should be simply considered as rough estimates.

Making use of $L_{\rm X}$ measured by \chandra\ and exploiting the information
provided by the ground-based optical observations, we can thus estimate the 
accretion rate of these AGNs in term of their Eddington ratio.
The results of these calculations are summarized in Table 4. Specifically, 
the values of the X-ray luminosity, unambiguously related to the AGN emission,
can be used to estimate the bolometric luminosity of the three naked AGNs, 
assuming a bolometric correction factor of 10 (e.g., Elvis et al. 1994). 
From the values of FWHM([O III])
we can derive an estimate for the stellar velocity dispersion (Greene \& Ho,
2005) and thus the black hole mass, using the $M_{\rm BH} - \sigma_{\star}$
correlation (Tremaine et al. 2002). We are now able to estimate the Eddington
ratio $L_{\rm bol}/L_{\rm  Edd}$ for the three sources. 

For Q2122-444 and Q2130-431 we find Eddington ratios of the order of 
$1\times 10^{-2}$, which is about one order
of magnitude larger than the threshold proposed by Nicastro et al. (2003) to 
separate AGNs with HBLRs from  non-HBLR sources.
On the other hand, the third source, Q2131-427, is in agreement with Nicastro 
et al. (2003) predictions. The large Eddington ratios found for Q2122-444 and 
Q2130-431 are more consistent with the findings of Bian \& Gu (2007), who, 
using a large sample of 90 Seyfert 2
galaxies, found that above a given $L_{\rm bol}/L_{\rm  Edd}$ threshold both 
HBLR and non-HBLR 
sources are present, whereas below that threshold only non-HBLR objects are 
found.

In summary, the X-ray observations of naked AGNs seem to indicate that AGNs 
without BLR are associated not only with low-luminosity AGNs, as suggested 
by Nicastro (2000) and Laor (2003), but also with fairly powerful AGNs, as 
hypothesized by Bian \& Gu (2007). This is also supported by the findings of
Wolter et al. (2005), who detected with \xmm\ three QSO 2 candidates, 
which do not show any sign of obscuration. These sources might be somewhat 
related to the naked AGNs discussed here, although their [O III]/H$\beta$ 
line ratios are more typical of NLS1 galaxies and they are one order of 
magnitude brighter in the X-ray band. 

Our results are also in agreement with the findings of 
Steffen et al. (2003), who, using Chandra deep field data, showed that a 
sizable fraction of the local AGN population (z=0.1--1) is without broad 
lines and spans a large range of X-ray luminosities: 
$L_{\rm X}\sim 10^{42} - 10^{44}{~\rm erg~s^{-1}}$.
Importantly, they show that AGNs without BLRs (comprising both obscured
objects and pure type 2 AGNs) in fact represent the dominant class 
of AGNs for X-ray luminosities below $\sim 4\times 10^{43}{~\rm erg~s^{-1}}$.

\section{Conclusions}
We have used data from snapshot \chandra\ observations to investigate the 
basic X-ray properties of three naked AGNs, in preparation of deeper X-ray
observations. The main results can be summarized as follows:

\begin{itemize}

\item The three AGNs are easily detected in the 0.3--8 keV band, despite the
short exposures.

\item The sources appear to be point-like objects at the \chandra\ ACIS-S
spatial resolution. However, the presence of weak extended emission cannot
be ruled out. No nearby sources are detected within a radius of 1\arcmin.

\item At first order, their X-ray spectral properties seem to be
consistent with those of Seyfert 1 galaxies, with typical photon 
indices of $\sim$1.8 and without strong intrinsic absorption.

\item The three sources are fairly bright in the
X-rays, with $L_{\rm 2-10 ~keV}$ of the order of few units in $10^{43}$ \lum, 
which are comparable with the X-ray luminosities observed in Seyfert 1 
galaxies (see Nandra et al. 1997) and larger than typical Seyfert 2 
luminosities (Turner et al. 1997).
\end{itemize}

The high detection rate ($>10\%$) of naked AGNs in the starting sample of 
Seyfert-like objects,
combined with the concrete possibility that they may be representative of
the large class of AGNs without BLRs, bolster the need 
to understand the nature of their central engine. A thorough investigation 
of the X-ray properties of the naked AGNs is crucial to reach this goal and 
to derive a more efficient way (compared to optical monitoring campaigns over 
periods of 25 years) to discover other members of this class, and shed new 
light on the AGN phenomenon.

\begin{acknowledgements} 
We thank the referee for his/her
comments and suggestions that improved the clarity of the paper.
MG acknowledges support by the Chandra Guest Investigator Program
under NASA grant 200997.  
\end{acknowledgements}

\begin{table}[ht]
\caption{ Source properties}
\begin{center}
\scriptsize
\begin{tabular}{lcccccccc}
\hline
\hline
\noalign{\smallskip}
 Name & RA  & DEC &  $ z$ &   $ N_{\rm H}^{\rm Gal}$ & FWHM(H$_\beta$)& [O III]/H$\beta$& $ m_B$ & $\alpha_{\rm OX}$ \\
\noalign{\smallskip}
     &( J2000)   & ( J2000)    &  & ($ 10^{ 20}{ \rm ~cm^{-2}}$) & $({\rm km~s^{-1}})$ & \\
\noalign{\smallskip}       
\hline
\noalign{\smallskip}
 Q2122-444 &$ 21^h26^m03.94^s$ & $ -44^o$  11' 19" &
 0.311 &  3.56 &  351 & 5.89 & 19.8 & -1.33\\
\noalign{\smallskip}
 Q2130-431 &$ 21^h33^m15.62^s$ & $ -42^o$  54' 24" &
 0.266 &  3.09 & 426 & 16.36 & 20.3 &-1.28\\
\noalign{\smallskip}
 Q2131-427 &$ 21^h34^m26.49^s$ & $ -42^o$  29' 56" &
 0.365  &  2.65 & 983 & 7.43& 20.5 &-1.31\\
\noalign{\smallskip}
\hline
\end{tabular}
\end{center}
\footnotesize
{\bf Columns Table 1}: 1= Source name. 2= Source coordinates. 3= Redshift.
4= Galactic column density. 5= Full width at half maximum for H$_\beta$
from Hawkins 2004 (no errors are reported). 5=[O III]$\lambda$5007/H$_\beta$
ratio from Hawkins 2004. 6= Average magnitude in the B band obtained from the
light curves provided by Hawkins 2004. For Q2122-444 the magnitude was obtained
from the all-sky catalogue at the ESO/ST-ECF Archive
produced by the US Naval Observatory (http://archive.eso.org/skycat/servers/usnoa), 
since no light curve was provided from Hawkins 2004. 6= value of
$\alpha_{\rm OX}$ derived from the best-fit
parameters of the $ \alpha_{\rm OX} - l_{\rm 2500\AA}$ relation
provided by Steffen et al. (2006).
\label{tab1}
\end{table}

\begin{table}
\caption{ Observation Log}
\begin{center}
\scriptsize
\begin{tabular}{lccccc}
\hline
\hline
\noalign{\smallskip}
 Name &  Date       & Exposure  & Frame time& Counts & Count Rate \\
\noalign{\smallskip}
    &    (yyyy/mm/dd) &( ks)    &  (s)      &     & (${\rm s^{-1}}$)\\
\noalign{\smallskip}       
\hline
\noalign{\smallskip}
 Q2122-444 & 2005/12/18 & 3.8 & 0.8 & $55\pm8$ & $ (1.4\pm0.2)\times10^{-2}$  \\
\noalign{\smallskip}
 Q2130-431 & 2005/12/14 &  2.9 & 0.4 &$79\pm10$ & $ (2.7\pm0.3)\times10^{-2}$ \\
\noalign{\smallskip}
 Q2131-427 & 2005/12/18 &  2.9 & 0.4 &$19\pm5$ & $ (6.6\pm1.5)\times10^{-3}$ \\
\noalign{\smallskip}
\hline
\end{tabular}
\end{center}
\footnotesize
{\bf Columns Table 2}: 1= Source name. 2= Date of observation. 3= Exposure time.
4= Frame time. 5= Number of counts extracted from a
circular region with a radius of 2''; the uncertainties are
calculated following Gehrels (1986). 6= Source count rate.
\label{tab2}
\end{table}

\begin{table}
\caption{ X-ray properties}
\scriptsize
\begin{center}
\begin{tabular}{lcccccc}
\hline
\hline
\noalign{\smallskip}
 Name  &  $ \Gamma $ &   $ HR $ & $ F_{\rm  0.5-8~keV}$ & $ L_{\rm  0.5-8~keV}$  & $ L_{\rm  2~keV}$ & $ L_{\rm  2~keV}^{\rm predicted}$\\
\noalign{\smallskip}
       &  &  (H-S)/(H+S)  & (${\rm erg~cm^{-2}~s^{-1}}$)& (${\rm 10^{43}~erg~s^{-1}}$)$^a$  & (${\rm10^{25}~ erg~s^{-1}~Hz^{-1}}$)  & (${\rm10^{25}~ erg~s^{-1}~Hz^{-1}}$)\\
\noalign{\smallskip}       
\hline
\noalign{\smallskip}
 Q2122-444 &
$ 1.8\pm0.4$  & $ -0.57\pm0.13$ & $ (1.0\pm0.2)\times10^{-13}$ 
& $ 3.2\pm0.4$& $ 2.0\pm0.5$ & $ 3.0$   \\
\noalign{\smallskip}
 Q2130-431  &  
$ 1.8\pm0.4$  & $ -0.39\pm0.12$ & $ (1.9\pm0.4)\times10^{-13}$ 
& $ 4.3\pm0.6$& $ 2.7\pm0.8$ & $ 1.8$ \\
\noalign{\smallskip}
 Q2131-427  &  
$2.1\pm0.9 $  & $ -0.60\pm0.27$ & $ (3.7\pm1.5)\times10^{-14}$ 
& $ 1.8\pm0.4$ & $ 1.2\pm0.8$& $ 2.5$\\
\noalign{\smallskip}
\hline
\end{tabular}
\end{center}
\footnotesize
{\bf Columns Table 3}: 1= Source name. 2= Photon index. 3= Hardness ratio.
4= Absorbed flux in the 0.5--8 keV band. 5= Intrinsic luminosity in the 0.5--8 
keV band. 6= Monochromatic luminosity at 2 keV obtained from \chandra\ observations.
7= Monochromatic luminosity at 2 keV estimated from the $ \alpha_{\rm OX}$ value given in Table 1.
\end{table}

\begin{table}
\caption{ Black hole masses and bolometric luminosities}
\scriptsize
\begin{center}
\begin{tabular}{lccccc}
\hline
\hline
\noalign{\smallskip}
 Name  &   FWHM([O III]) & $\sigma_{\star}$ &   $M_{\rm BH} $ & $ L_{\rm Edd}$   & $ L_{\rm bol}/L_{\rm  Edd}$ \\
\noalign{\smallskip}
       & $({\rm km~s^{-1}})$ & $({\rm km~s^{-1}})$  & ($M_\odot$)  & (${\rm erg~s^{-1}}$) \\
\noalign{\smallskip}       
\hline
\noalign{\smallskip}
 Q2122-444 & 714  &  227 & $ 2.2\times10^{8}$ & $ 2.8\times10^{46}$ &  $ 8.0\times10^{-3}$\\
\noalign{\smallskip}
Q2130-431  & 652  &  207 & $ 1.5\times10^{8}$ & $ 1.9\times10^{46}$ &  $ 1.5\times10^{-2}$\\
\noalign{\smallskip}
Q2131-427  & 1143 &  363 & $ 1.5\times10^{9}$ & $ 1.9\times10^{47}$ &  $ 5.1\times10^{-4}$\\
\noalign{\smallskip}
\hline
\end{tabular}
\end{center}
\footnotesize
{\bf Columns Table 4}: 1= Source name. 2= Full width at half maximum for
[O III]$\lambda$5007 (from Hawkins 2004). 3= Stellar velocity dispersion obtained using the formula $\sigma_{\star}=FWHM([O~III])/2.35/1.34$ (Greene \& Ho 2005).
4= Black hole mass derived from $M_{\rm BH}=10^{8.13}[\sigma_{\star}/(200~{\rm km~s^{-1}})]^{4.02}~M_\odot$ (Tremaine et al. 2002). 5= Eddington luminosity.
7= Bolometric luminosity normalized to the Eddington value; the bolometric luminosity 
was obtained by multiplying $L_{\rm 2-10 ~keV}$ by a factor of 10.
\end{table}


\begin{thebibliography}{}
\bibitem[Antonucci 1993]{anto93} Antonucci, R. 1993, ARA\&A, 31, 473

\bibitem[Arnaud 1996]{arn96} Arnaud, K. 1996, in ASP Conf. Ser. 101, 
Astronomical Data Analysis Software and Systems V, ed. G. Jacoby \& J. 
Barnes (San Francisco: ASP), 17

\bibitem[Bauer et al. 2004]{bau04} Bauer, F.E., et al. 2004, AJ, 128, 2048

\bibitem[Bennet et al. 2003]{ben03} Bennet, C.L. et al. 2003, ApJS, 148, 1

\bibitem[Bian \& Gu 2006]{bian06} Bian, W. \& Gu. Q, 2006, ApJ, 657, 159

\bibitem[Dorman \& Arnaud 2001]{dorm01} Dorman, B. \& Arnaud, K.A. 2001 in
Astronomical Data Analysis Software and Systems X, ASP Conference
Proceedings, Vol. 238. Edited by F. R. Harnden, Jr., Francis A. Primini,
and Harry E. Payne. San Francisco: Astronomical Society of the Pacific,
p.415

\bibitem[Cappi et al. 2006]{cappi06} Cappi, M., et al. 2006, A\&A, 446, 459

\bibitem[Cash 1979]{cash79} Cash, W. 1979, ApJ 228, 939

\bibitem[Donato et al. 2005]{don05} Donato, D., Sambruna, R.M, \& Gliozzi, M. 2005,
A\&A, 433, 1163

\bibitem[Dudik et al. 2005]{dudi05} Dudik, R.P., et al. 2005, ApJ, 620, 113

\bibitem[Elvis et al. 1994]{elv94} Elvis, M. et al. 1994, ApJS, 95, 1

\bibitem[Gehrels]{gehr86} Gehrels, N. 1986, ApJ, 303, 336

\bibitem[Georgantopoulos \& Zezas 2003]{geo03} Georgantopoulos, I. \& Zesas, 
A. 2003, ApJ, 594, 704

\bibitem[Gliozzi et al. 2003]{glio03} Gliozzi, M., Sambruna, R.M., Brandt, 
W.N. 2003, A\&A, 408, 949

\bibitem[Gliozzi et al. 2004]{glio04} Gliozzi, M., Sambruna, R.M., Brandt, 
W.N., Mushotzky, R., Eracleous, M. 2004, A\&A, 413, 139

\bibitem[Gliozzi et al. 2006]{glio06} Gliozzi, M., Papadakis, I.E. \& Brinkmann, W.P., 2007,
ApJ, 656, 691


\bibitem[Greene \& Ho 2005]{gree05} Greene, J.E. \& Ho, L.C. 2005, ApJ, 627, 721

\bibitem[Gu \& Huang 2002]{gu02} Gu, Q. \& Huang, J. 2002, ApJ, 579, 205

\bibitem[Hawkins 2004]{hawk04} Hawkins, M.R.S. 2004, A\&A, 424, 519

\bibitem[Laor 2003]{lao03} Laor, A. 2003, ApJ, 590, 86

\bibitem[Maiolino \& Risaliti 2007]{maio07} Maiolino R., \& Risaliti, G. 2007, astro-ph/0701109

\bibitem[Marshall et al. 2004]{marsh04} Marshall, H.L. 2005, ApJS, 156, 13

\bibitem[Nandra et al. 1997]{nand97} Nandra, K., George, I.M., Mushotzky, R.F., Turner, T.J., Yaqoob, T.
1997, ApJ, 477, 602

\bibitem[Netzer \& Trakhtenbrot 2007]{netz07} Netzer, H. \& Trakhtenbrot, B. 2007,
ApJ, 654, 754

\bibitem[Nicastro 2000]{nic00} Nicastro, F. 2000, ApJ, 530, L65

\bibitem[Nicastro et al. 2003]{nic03} Nicastro, F., Martocchia, A.,
\& Matt, G. 2003, ApJ, 589, L13

\bibitem[Pogge 2000]{pogg00} Pogge, R.W. 2000, New Astronomy Review 44, 381

\bibitem[Raiteri et al. 2006]{rait06} Raiteri, C., et al. 2006, A\&A, 459, 731

\bibitem[Sambruna et al. 2002]{sambr02} Sambruna, R.M., et al. 2002, ApJ, 571, 206

\bibitem[Sbarufatti et al. 2006]{sbarr06} Sbarufatti, B., Falomo, R., Treves, A., 
Kotilainen, J. 2006, A\&A, 457, 35
 
\bibitem[Steffen et al. 2003]{stef03} Steffen, A.T., Barger, A.J., Cowie, L.
L., Mushotzky, R.F., Yang, Y. 2003, ApJ, 596, L23

\bibitem[Steffen et al. 2006]{stef06} Steffen, A.T., et al. 2006, AJ, 131, 2826

\bibitem[Tran 2001]{tran01} Tran, H.D. 2001, ApJ, 554, L19

\bibitem[Tremaine et al. 2002]{trem02} Tremaine, S., Gebhardt, K., Bender, R., et al. 2002, ApJ, 574, 740

\bibitem[Turner et al. 1997]{turn97} Turner, T.J., George, I.M., Nandra, K., \& Mushotzky, R.F., 1997,
ApJ, 113, 23

\bibitem[Urry \& Padovani 1995]{urr95} Urry, M. \& Padovani, P. 1995, PASP, 107
 
\bibitem[Wolter et al. 2005]{wolt05} Wolter A., Gioia, I.M., Henry, J.P., 
Mullis, C.R. 2005, A\&A 444, 165
\end{thebibliography}
\end{document}